\documentclass[aps,prb,twocolumn,floatfix]{revtex4-1}

\usepackage{graphicx}
\usepackage{amsmath}
\usepackage{amsfonts}
\usepackage{amssymb}
\usepackage{dcolumn}
\usepackage{braket}
\usepackage{commath}
\usepackage{epstopdf}
\usepackage{hyperref}

\graphicspath{{figs_ibrahim/}}

\usepackage{color} 

\newcommand{\red}{\textcolor{black}}

\usepackage{ulem}   

\begin{document}

\title{\red{Double well potentials with a quantum moat barrier
or a quantum wall barrier give rise to similar entangled
wave functions}}

\author{A. Ibrahim}

\author{F.~Marsiglio}
\email{fm3@ualberta.ca}

\affiliation{Department of Physics, University of Alberta, Edmonton, Alberta, Canada, T6G~2E1}

\begin{abstract}
The solution to a problem in quantum mechanics is generally a linear superposition of states.
\red{The solutions for double well potentials epitomize this property, and go even further than this: they can often be 
described by an effective model
whose low energy features can be described by two states --- one in which the particle is on one side of the barrier, and a second where
the particle is on the other side. Then the ground state remains a linear superposition of these two macroscopic-like states. In this paper
w}e illustrate that this \red{property is achieved similarly} with \red{an attractive potential}
that separates two regions of space\red{, as opposed to the traditionally repulsive one}. In explaining how this comes about we revisit the concept of ``orthogonalized plane waves,''
first discussed in 1940 to understand electronic band structure in solids, along with the accompanying concept of
a pseudopotential. We show how these ideas manifest themselves in a simple double well potential, whose ``barrier''
consists of a moat instead of the conventional wall.
\end{abstract}

\date{\today}
\maketitle

\section{introduction}

There are many fascinating aspects of quantum mechanics which clearly disturb our classical intuition. Tunnelling is one example, and leads
directly to the notion of state superposition (and therefore entanglement) in the case of a double well potential.\cite{textbooks} 
A double well potential can take on
many forms, but the essence is captured in the one-dimensional example of two spatial regions consisting of wells, separated by
a potential barrier. In general the energy of a particle in this system is lowered by bringing these two regions closer together --- this is the
essence of molecular bonding in a diatomic molecule. Tunnelling allows the particle to be ``shared'' between these two regions; in
quantum mechanics one says that the ground state of this system is best described as a linear superposition of two states, one in which
the particle is in the left well region, and one in which it is in the right well region.

The subject of this paper is to address the question: to what extent does the separation into two regions have to be effected by a barrier
that consists of a wall. \red{That is, in order to achieve macroscopic-like states (particle on the left side vs. particle on the right side), can
we use a potential other than a wall to divide the space into two distinct regions?}
\red{For the remainder of this paper, the term ``moat'' will refer to a negative attractive potential, while ``wall'' will refer to a positive repulsive potential.}
Even in medieval times castle-builders knew that 
a moat\cite{remark_page} 
could serve as an 
effective `barrier' for a castle.\cite{remark2} In quantum mechanics as well, most students know that a well can be an effective
deterrent for transmission. In fact, for a $\delta$-function potential, the reflectance coefficient for an incoming plane wave with
energy $E$ is given by
\begin{equation}
|R|^2 = {1 \over 1 + {2\hbar^2 E \over m\alpha^2}}
\label{reflection}
\end{equation}
where $\alpha$ is the strength (in units of energy*length) of the $\delta$-function potential. This expression makes it clear
that the reflectance approaches unity, i.e. the particle completely bounces back, as $|\alpha | \rightarrow \infty$ for both
the wall ($\alpha > 0$) and the moat ($\alpha < 0$).\red{\cite{remark2b}}

In this paper we will address this question with the simplest double well potential --- an infinite square well divided into two regions
with a $\delta$-function potential in the middle. We make this choice because everything is known analytically,\cite{remark3} and
comparisons between different scenarios can readily be made. We start with a historical context. This involves a much more complicated
problem (metallic behaviour in an infinite array of atoms forming a three-dimensional solid), but the current example 
will explain the conceptual advance made in that instance, but in a much simpler context.
The exact solution is trivial, but we then explain how one would proceed without such a solution (as was the situation
historically). In this way we can illustrate the difficulty with a ``brute force'' approach, and implement the remedy 
to the more complicated problem on a much simpler one.
This will introduce the novice to the notion
of ``orthogonalization'' and the pseudopotential. The goal is to make it clear how a potential moat ends up behaving like a potential
wall, i.e. both act as an effective barrier.

\section{Historical Context}

\label{historical}

Soon after the advent of quantum mechanics, physicists wanted to understand the behaviour of electrons in a solid.
At the time a big leap occurred with Bloch's Theorem;\cite{bloch29} this allowed the formulation of the one-electron
problem in a periodic potential to be simplified to a single unit cell in the solid, and introduced a quantum ``number'' $k$ 
(really a wave vector).\cite{ashcroft76,grosso00} This was followed by a series of advances towards determining the
electronic band structure for a single electron in a solid, since the periodicity, exploited explicitly 
in the Kronig-Penney model,\cite{kronig31}
resulted in an energy dispersion [i.e. $E(k)$] that would determine the transport (and other) properties of the material. Wigner and
Seitz then developed the so-called cellular method,\cite{wigner33} and further developments utilized various forms of plane-wave
expansions.\cite{slater34}

The advance with which the present paper is concerned was by Herring, in 1940.\cite{herring40} It is called the ``Orthogonalized
Plane-Wave (OPW) method.'' The recognized difficulty at the time was that the plane wave expansion is extremely inefficient
at obtaining the deeply bound states, i.e. the so-called ``core'' states of the atom. These are states that are not
significantly modified by the presence of an array of atoms. In other words, using a specific example like sodium, the $1s$ state,
as it exists with of order Avogadro's number of sodium atoms periodically arranged in a solid, is hardly different from what one
would calculate for a single sodium atom. Moreover, the more relevant states with significant amplitude in the region between
the cores --- the so-called interstitial region (e.g. the $3s$ state in sodium), would also
require an inefficiently large number of plane-wave states for an accurate calculation. 
Herring recognized that this inefficiency stemmed from the state's behaviour in the atomic core regions; far more plane waves are needed to accurately generate oscillations in the core regions compared to the interstitial regions.
On the other hand, this oscillatory
behaviour was not really important, as the transport properties of the material are largely determined by the behaviour of the
$3s$ state in the interstitial region, i.e. between the cores.

Since these core states are approximately already known, as argued above, then he could introduce an explicit 
orthogonalization of the relevant states to these core states. 
This establishes a new basis, of modified plane waves, or, as they are more descriptively
called, orthogonal plane waves. Further insight was added later,\cite{phillips59,antoncik59} in that the original problem
could be reformulated as one involving only these modified plane wave-like states, but with an effective
Hamiltonian. This new Hamiltonian consisted
of both the original periodic potential, and a periodic induced ``potential'', due to the existence of the core states and the
required orthogonalization. Furthermore, it is straightforward to show that this induced ``potential'' would make a positive (i.e. repulsive)
contribution to the one-particle potential, and would therefore cancel to some extent the negative (i.e. attractive) potential that
is a priori present in a solid, due to the positively charged atomic cores. The sum of these two potentials, 
called the pseudopotential, would
therefore be much smaller than originally surmised from the atomic cores alone, and this would explain why the free electron-like
description was so successful. This fact also increases the efficiency of the new, orthogonalized, basis states, for describing
the interstitial regions accurately.

The problem in the present paper is much simpler: there is only one core, and for the present it is given by an attractive
$\delta$-function potential in the center of an infinite square well. If this potential were positive, it would divide the infinite
square well into two regions, and the \sout{particle} \red{solution} would have all the characteristics of a particle in a double well potential.
In the next section we review this solution,\cite{remark3,textbooks} and illustrate how a strong moat is as effective a barrier as
a strong wall.

\section{Equivalence of the strongly repulsive and attractive $\delta$-function potential}

\begin{figure}[h]
\includegraphics[width=8.5cm]{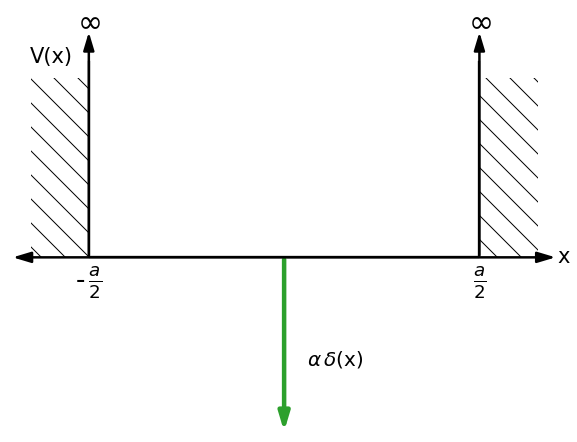}
\caption{Does this look like a double well potential? A schematic of the $\delta$-function potential, shown as a ``moat'' 
(i.e. $\alpha < 0$) inside the infinite square well.
For $\alpha >0$ it would point upwards and represent a ``wall'' between the left side and the right side of this infinite
square well. The purpose of this paper is to convince the reader that the moat behaves very much like the wall, and indeed,
this figure represents a double well potential.}
\label{Fig1}
\end{figure}
In Fig.~\ref{Fig1} we provide a schematic of the potential used in this paper. With a definition of $k \equiv \sqrt{2mE/\hbar^2}$,
where $E$ is the energy, a very straightforward\red{\cite{remark4}} analysis leads to the
scattering\red{\cite{remark5}} solutions,
\begin{eqnarray}
\psi_e(x) & = & A_{k_e} \ \sin{k_e ({a \over 2} - |x|) } \phantom{aaaaaaaaaaaaa}   {\rm (even)}\\
\psi_o(x) & = & A_{k_o} \ {\rm sgn}(x) \ \sin{k_o ({a \over 2} - |x|)}   \phantom{aaaaaaaa} {\rm (odd)} \\
A_k & = & \sqrt{2 \over a} {1 \over \sqrt{1 - {\sin{ka} \over ka}}} \phantom{aaaaaaaaaaaaaaaa} {\rm (either)},
\label{wavefn}
\end{eqnarray}
where the subscripts `$e$' and `$o$' signify `even' and `odd' solutions, respectively, and $k$ in the definition of energy and in 
the last equation refers
to either $k = k_e$ or $k = k_o$. These are scattering states with respect to the $\delta$-function potential, but the presence
of the infinite square well gives rise to the quantization conditions [$z_0 \equiv m\alpha a/(2 \hbar^2$)]:
\begin{eqnarray}
\tan{k_ea \over 2} & = & -{1 \over z_0}\left({k_ea \over 2}\right)  \phantom{aaaaaaaaaaaa}   {\rm (even)} \label{even_eig} \\
k_oa/2 & = & n \pi, \ \ \ n=1,2,3...  \phantom{aaaaaaaa} {\rm (odd)}.
\label{odd_eig}
\end{eqnarray}
\begin{figure}[h]
\includegraphics[width=5.5cm,angle = -90]{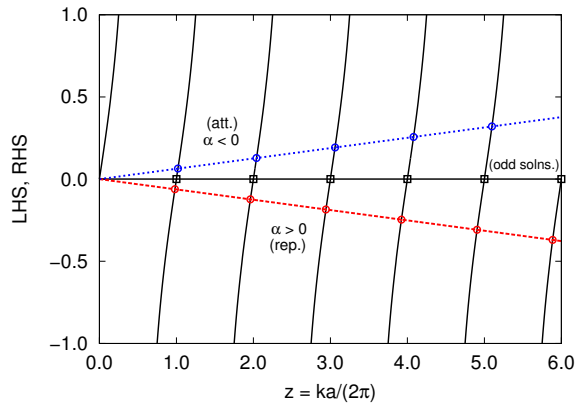}
\caption{\red{We plot the} Left-Hand-Side (LHS) (solid black curve with the characteristic branches of the $tan$ function) and the 
Right-Hand-Side (RHS) of Eq.~(\ref{even_eig}) (i.e. the even case) vs. $z \equiv ka/(2\pi)$ \red{to show the solutions graphically}. 
The RHS depends on the strength of
the $\delta$-function potential, $\alpha$. In all cases the RHS is linear in $z$ with a zero intercept, and has a negative
(positive) slope for the repulsive, $\alpha > 0$ (attractive, $\alpha < 0$) case. Two example cases are shown, the
red dashed curve with $z_0 \equiv m\alpha a/(2\hbar^2) = 50$ (repulsive, $\alpha > 0$), and the
blue dotted curve with $z_0 =-50$ (attractive, $\alpha < 0$). The solutions are shown by circles where these respective
lines intercept the $tan$ function. The solutions for the odd case are trivial, and are simply given by $ka = 2\pi n$, $n=1,2,3,...$,
and are indicated by the squares along the $z$-axis.
}
\label{Fig2}
\end{figure}
As expected the odd solutions are impervious to the $\delta$-function potential, and remain as they were [Eq.~(\ref{odd_eig})]
in the absence of this potential. 
The even solutions
have energies determined by Eq.~(\ref{even_eig}). Furthermore, a bound state may exist [if $\alpha < -4a \ \hbar^2/(2ma^2)$].
In that case the solution is
\begin{eqnarray}
\psi_b(x) & = & A_b \ \sinh{\kappa({a \over 2} - |x|)} \phantom{aaaaaaaaaa}   {\rm (bound)}\\
A_b & = & \sqrt{2 \over a} {1 \over \sqrt{{\sinh{\kappa a} \over \kappa a} - 1}} \phantom{aaaaaaaaaaa} {\rm (bound)},
\label{wavefnb}
\end{eqnarray}
where $\kappa \equiv \sqrt{-2mE_b/\hbar^2}$ and the subscript `b' stands for `bound'. We should emphasize that we are not interested
in the bound state per se, but in the scattering states, and particularly the two lowest ones. In Fig.~\ref{Fig2} we show a graphical
construction of Eq.~(\ref{even_eig}) to determine the energies; details concerning this plot are in the figure caption. It is clear that, as
expected, for the repulsive case the solutions tend to come in pairs (since we have chosen $\alpha$ to be large), and they
alternate between even and odd solutions (with the even solution slightly lower in energy, than the odd solutions). The even solution
has a slightly lower energy than it would have if it were isolated in an infinite well of width $a/2$, confirming the familiar 
physics\cite{textbooks,remark3} that the ability to tunnel to the neighboring region lowers the overall energy. These two solutions have
probability density equally distributed in the two regions of the infinite square well, and represent the ``bonding'' and ``anti-bonding''
solutions to this problem.

\begin{figure}[h]
\includegraphics[width=3.7cm,angle = -90]{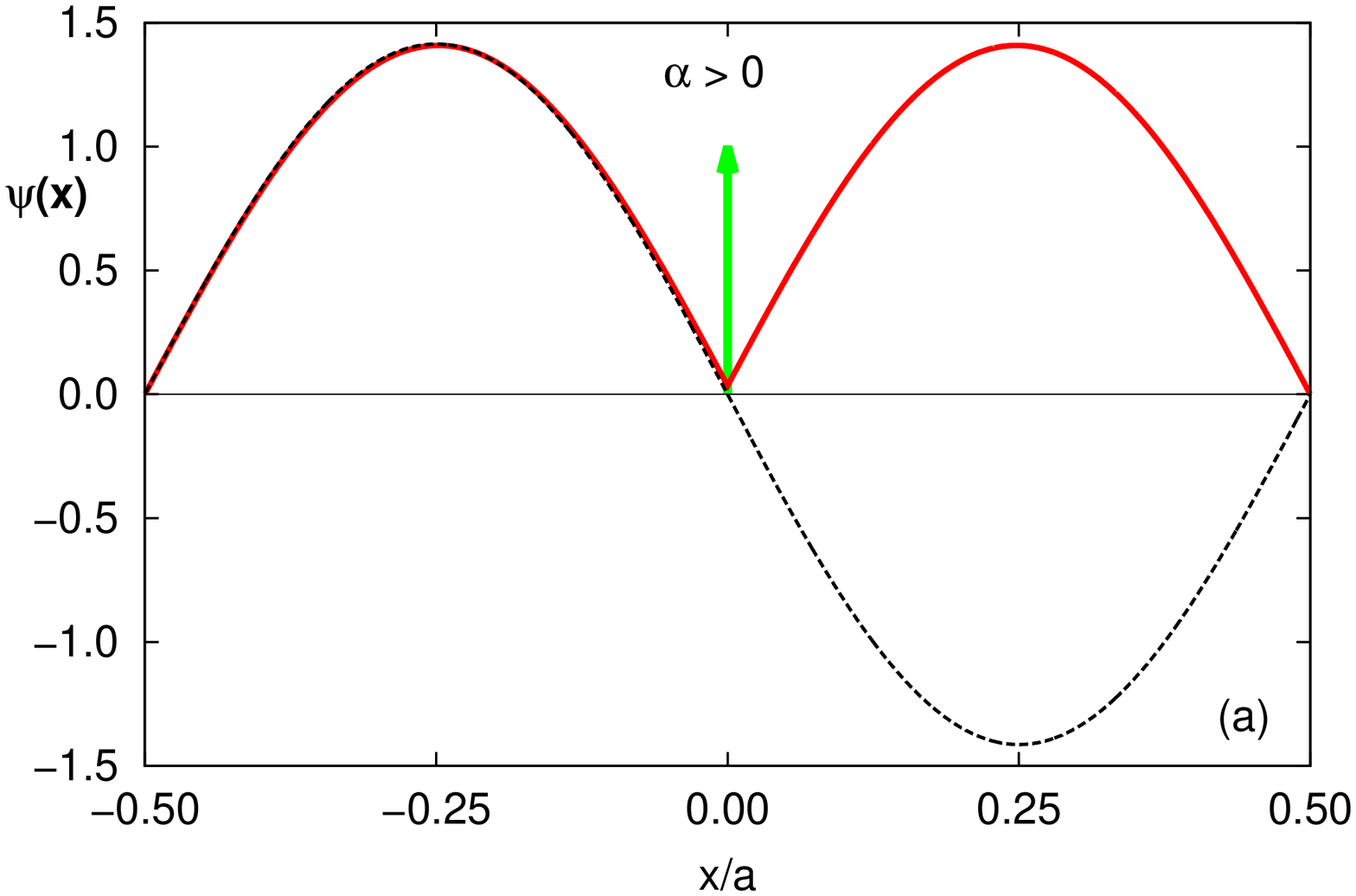}
\includegraphics[width=3.7cm,angle = -90]{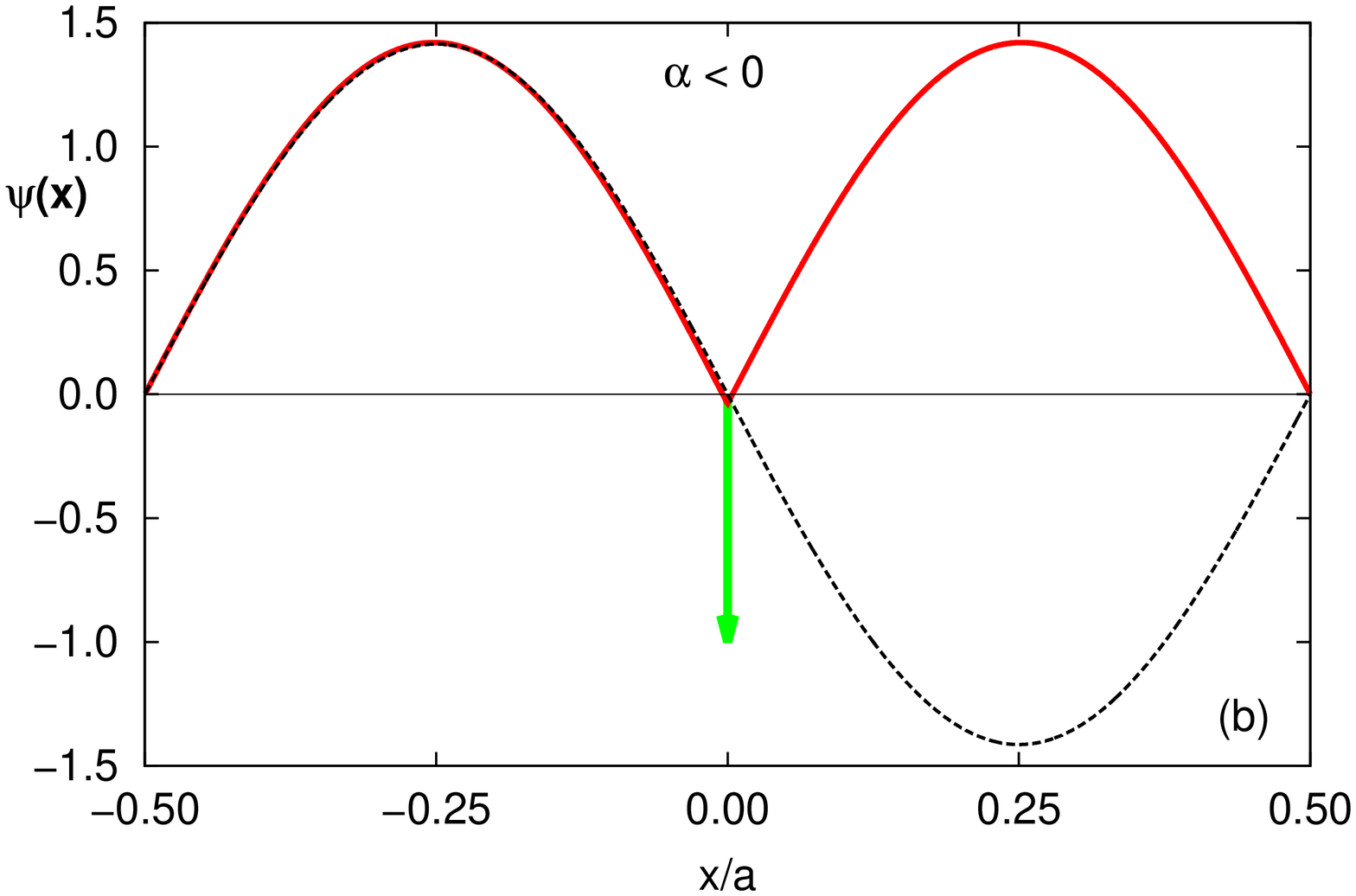}
\includegraphics[width=3.7cm,angle = -90]{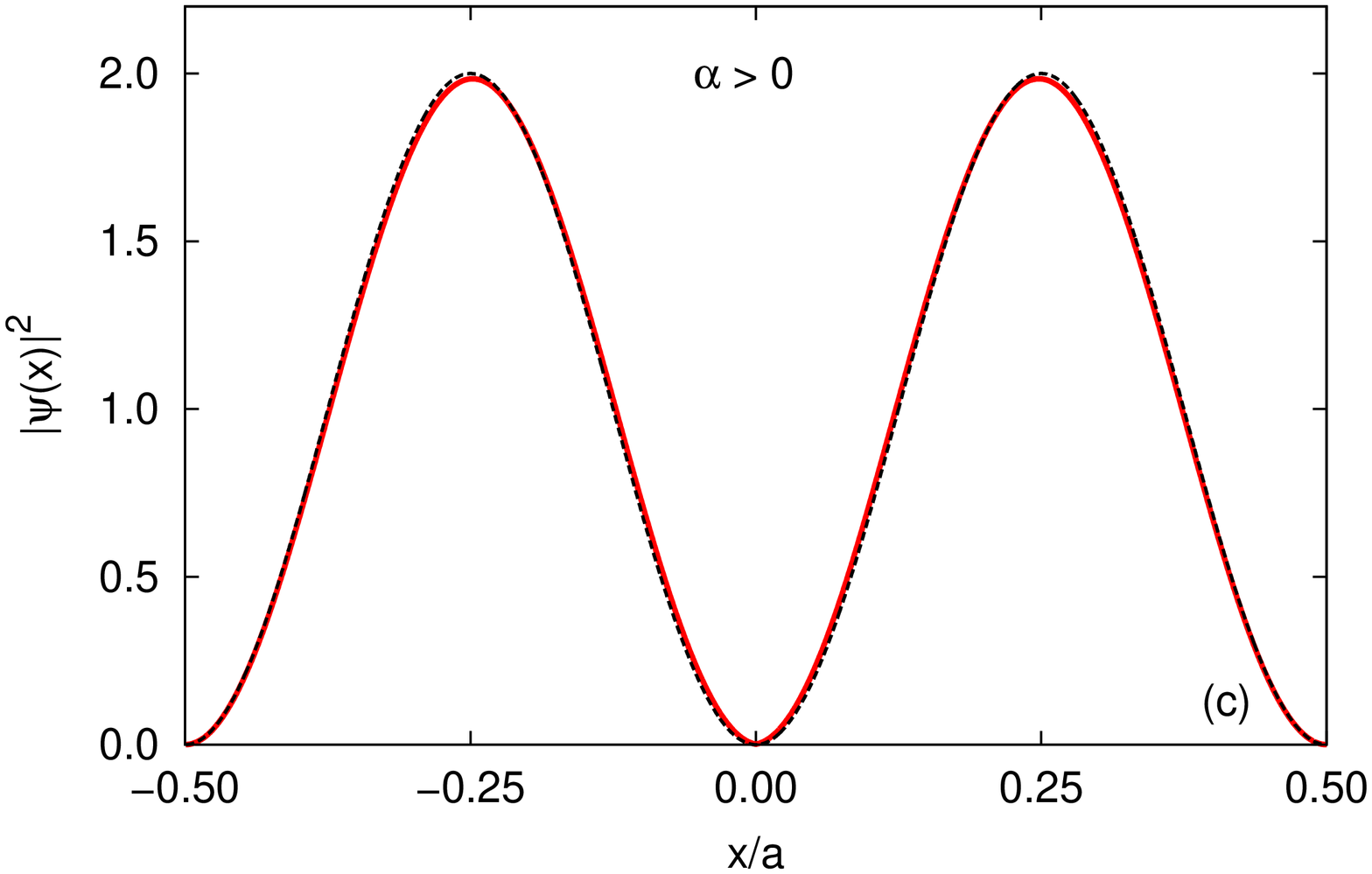}
\includegraphics[width=3.7cm,angle = -90]{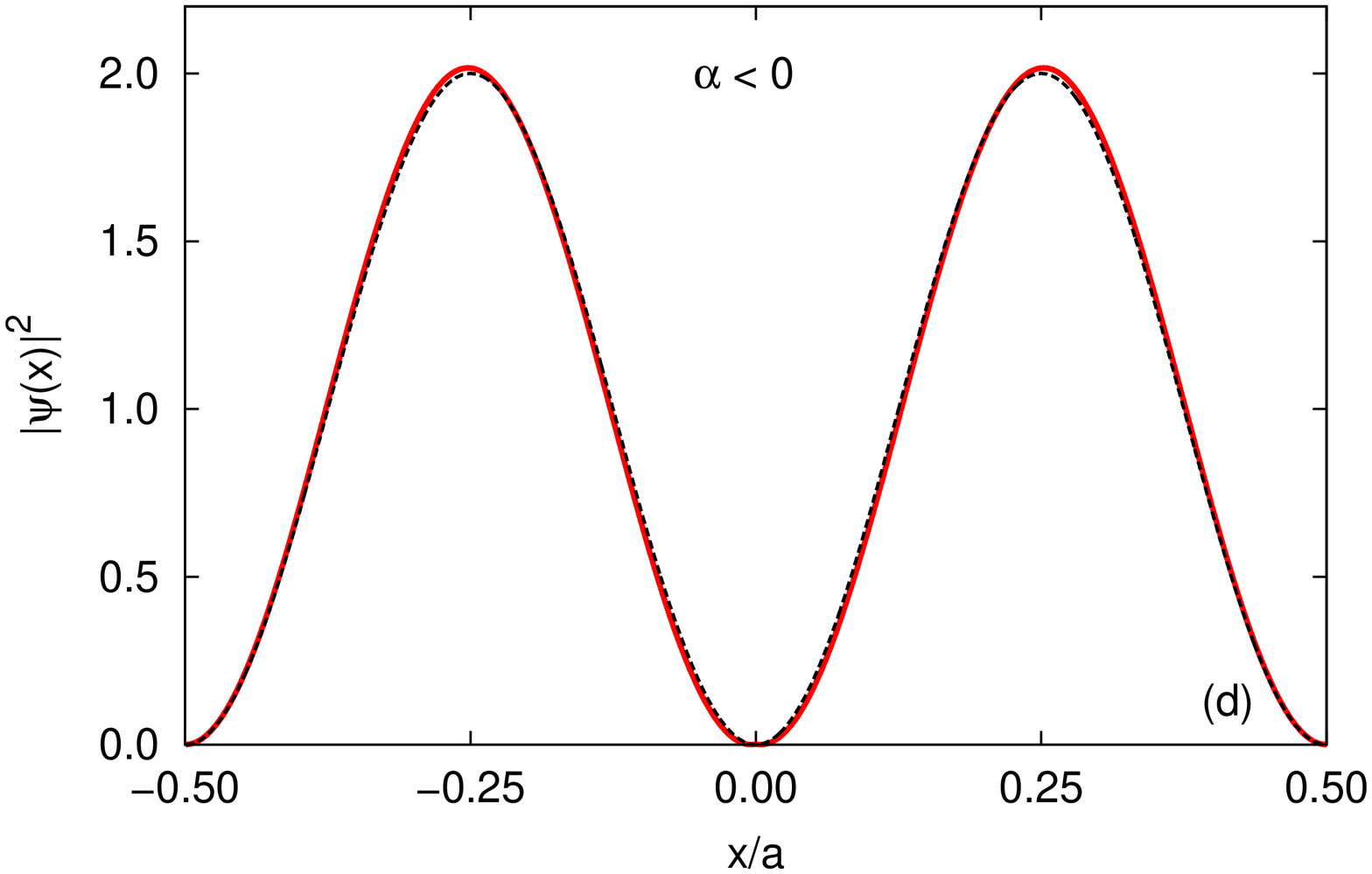}
\caption{The two lowest energy scattering states: their wave functions (top panels) and their probabilities (bottom panels) 
for a large value of $|\alpha| = 50  a\  \pi^2 \hbar^2/(2ma^2)$ for $\alpha > 0$  (left side) and $\alpha < 0$ (right side) as a function of position. Even wave
functions are shown with a solid (red) curve, while odd wave functions are shown with a dashed (black) curve. Their probability
amplitudes are barely distinguishable from one another.
The left and right sides also look nearly the same, indicating that when the strength is high enough, a
$\delta$-function moat is as effective \red{a barrier} as a $\delta$-function wall. Close inspection of the bottom two panels indicates that the
moat actually keeps more wave function amplitude away from the central ``barrier'' region than the wall does (see Fig.~\ref{Fig4} below).
The green vertical arrow is a schematic indication of the potential involved (repulsive on the left, and attractive on the right).}
\label{Fig3}
\end{figure}
In addition, however, for the attractive case the solutions also come in pairs, this time with the odd solution having slightly lower energy.
So even though the potential illustrated in Fig.~\ref{Fig1} with an attractive $\delta$-function potential does not at all resemble a
double well potential, it behaves like one, as far as the energies are concerned. To confirm this is the case we show the
wave functions and probabilities in Fig.~\ref{Fig3}, first for the repulsive case (two left-side panels) and then for the attractive case
(two right-side panels). It is clear that the $\delta$-function moat behaves much in the same way as the $\delta$-function wall. We explain why in the next section.

\section{Orthogonalized Eigenstates}

This problem is sufficiently simple that we have an exact solution --- this is not normally the case.
If we were to use a more arbitrary potential in lieu of
the $\delta$-function, then we would in general have to resort to a numerical
solution. The numerical procedure is straightforward, and was demonstrated in Ref.~[\onlinecite{marsiglio09}] for such a
general potential, and more specifically in Ref.~[\onlinecite{remark3}] and by Jelic et al. in Ref.~[\onlinecite{textbooks}]
for double well potentials
similar to those discussed here. In the numerical work we first shift the infinite square well to have a domain $0 < x < a$
so that the internal potential is centred at $x= a/2$ instead of $x=0$. This then allows for a convenient basis, consisting of the
eigenstates of the infinite square well, i.e.
\begin{equation}
\phi_j(x) = \sqrt{2 \over a} \sin{\left({\pi j x \over a}\right)}, \phantom{aaaaa} j = 1,2,3....
\label{basis}
\end{equation}
In fact, precisely the $\delta$-function potential depicted in Fig.~\ref{Fig1} was solved in this way in Ref.~[\onlinecite{marsiglio09}].
That paper's Fig.~4 illustrates that convergence using this method becomes difficult as a square well becomes narrower
and narrower and approaches a $\delta$-function potential, as considered in the present paper. Why is this so, particularly in the case
of the $\delta$-function moat? The reason is as follows.
The $\delta$-function moat will support a bound state for a sufficiently large $|\alpha|$; this result is analytically known, and
the wave function, not surprisingly, is sharply peaked and even contains a cusp. To properly describe such a function with
the basis set Eq.~(\ref{basis}) is a known difficult problem, and one needs to resort to higher and higher Fourier components
to get this right. We are interested in the scattering states, but the scattering states need to be orthogonal to the bound state
(as well as to one another). This is the same problem faced by Herring\cite{herring40} with a periodic array of atomic cores and
the bound states associated with these cores. 
We thus proceed as he did, and define a new basis, which is automatically
orthogonal to the bound state. We write a new basis set,
\begin{equation}
|\tilde{\phi}_j \rangle  = |\phi_j \rangle - |\phi_b \rangle \langle \phi_b | \phi_j \rangle,
\label{basis_new}
\end{equation} 
where $|\phi_b \rangle $ is the normalized bound state and $\langle x | \phi_j \rangle $ is just the basis state of Eq.~(\ref{basis}).
It is easy to verify that $\langle \phi_b | \tilde{\phi}_j \rangle = 0$ so they are indeed orthonormal. Substituting this into the
Schr\"odinger equation,
\begin{equation}
(\hat{H}_0 + V)|\psi \rangle = E |\psi \rangle,
\label{schrod}
\end{equation}
we now proceed to expand the desired wave function in terms of the $|\tilde{\phi}_j \rangle$ basis, 
\begin{equation}
|\psi\rangle = \sum_j c_j |\tilde{\phi}_j \rangle,
\label{expand}
\end{equation}
where the $c_j$ are the (unknown) coefficients.
Taking the inner product with $\langle  \tilde{\phi}_i |$ and substituting Eq.~(\ref{basis_new}) leads to
\begin{equation}
\sum_j \langle \phi_i | \left(\hat{H}_0 + V + (E - E_b)| \phi_b \rangle \langle \phi_b | \right) |\phi_j \rangle c_j = E c_i.
\label{mat1}
\end{equation}
This is now a matrix equation for a very different Hamiltonian, with a so-called pseudopotential,
\begin{equation}
V_{\rm ps} \equiv V + (E - E_b)| \phi_b \rangle \langle \phi_b |.
\label{pseudo}
\end{equation}
This is indeed a peculiar potential, in that it is non-local \red{(see below in Eq.~(\ref{coord}) for a concrete realization of the
non-locality of this potential)} and also depends on the eigenvalue $E$ that we are seeking.
As explained in Sec.~\ref{historical}, the pseudopotential interpretation of the orthogonal wave expansion came many years later.\cite{phillips59,antoncik59} Equation~(\ref{mat1}) is a secular equation, and can be solved by exact diagonalization, with
matrix elements being the sum of the three separate components
\begin{equation}
\langle \phi_i | \hat{H}_0  |\phi_j \rangle, \phantom{a} \langle \phi_i | V  |\phi_j \rangle, \phantom{a} {\rm and}
 \phantom{a} (E-E_b) \langle \phi_i | \phi_b \rangle \langle \phi_b | \phi_j \rangle .
\label{mat_elements}
\end{equation}
Now, because we know everything about this problem, including the exact bound state (with account for the infinite square well),
we can use these solutions in the third component of the matrix elements listed in (\ref{mat_elements}). We can insert
these matrix elements into Eq.~(\ref{mat1}), diagonalize the resulting Hamiltonian
and find the eigenvalues $E$ (this will require iteration, because $V_{\rm ps}$
depends on $E$) and the eigenvector components, $c_j$. The wave function can then be constructed from these eigenvector
components using Eq.~(\ref{expand}) and
Eq.~(\ref{basis_new}). This wave function has to be explicitly normalized (since, for an orthonormal set $\{ |\phi_j \rangle \}$, 
Eq.~(\ref{basis_new}) shows that the $\{ |\tilde{\phi}_j \rangle \}$ set will not be normalized). We have confirmed that the result
of this exercise for the scattering states agrees exactly with the analytical result (or with the converged numerical 
result before orthogonalizing to the bound state). 

But the true test of using an orthogonal plane wave expansion with the ensuing pseudopotential is if we use the
approximate bound state wave functions in Eq.~(\ref{mat_elements}). This would be like using the core
state wave functions as obtained in the tight-binding approximation for the problem facing Herring,\cite{herring40}
or, in our simple example, the bound state wave function for an isolated $\delta$-function potential at the origin. Contrary to
the exact solution given in Eq.~(\ref{wavefnb}), for the isolated potential the wave function is
\begin{equation}
\phi_b(x) = \sqrt{\kappa}e^{-\kappa |x|}, \phantom{aaaaa} {\rm with } \phantom{aaa} E_b = -{1 \over 2} {m \alpha^2 \over \hbar^2},
\label{app_wavefnb}
\end{equation}
where $\kappa = -m \alpha/\hbar^2$. Since $\kappa \equiv \sqrt{-2mE_b/\hbar^2}$ we obtain $E_b$ as written in Eq.~(\ref{app_wavefnb}).
Since we have now shifted the infinite square well, we will replace $x \rightarrow (x-a/2)$ in Eq.~(\ref{app_wavefnb}).
In the third matrix element of Eq.~(\ref{mat_elements}) we require
\begin{eqnarray}
\langle \phi_b | \phi_j \rangle &=& \int_0^a \ dx \langle \phi_b | x \rangle \langle x | \phi_j \rangle \nonumber \\
 & = &  \int_0^a \ dx \phi_b^\ast(x) \phi_j(x) \nonumber \\
 &=& \int_0^a \ dx \ \sqrt{\kappa} e^{-\kappa |x-a/2|} \sqrt{2 \over a} \sin{\left({j \pi x \over a}\right) } \nonumber \\
 & = & {2 \sqrt{2 \kappa a} \over (\kappa a)^2 + (j\pi)^2}\left( \kappa a \sin{\left({j \pi \over 2}\right)} + j\pi e^{-{\kappa a \over 2}}\right).
 \label{3rd_mat_ele}
\end{eqnarray}
\begin{figure}[h]
\includegraphics[width=5.0cm,angle = -90]{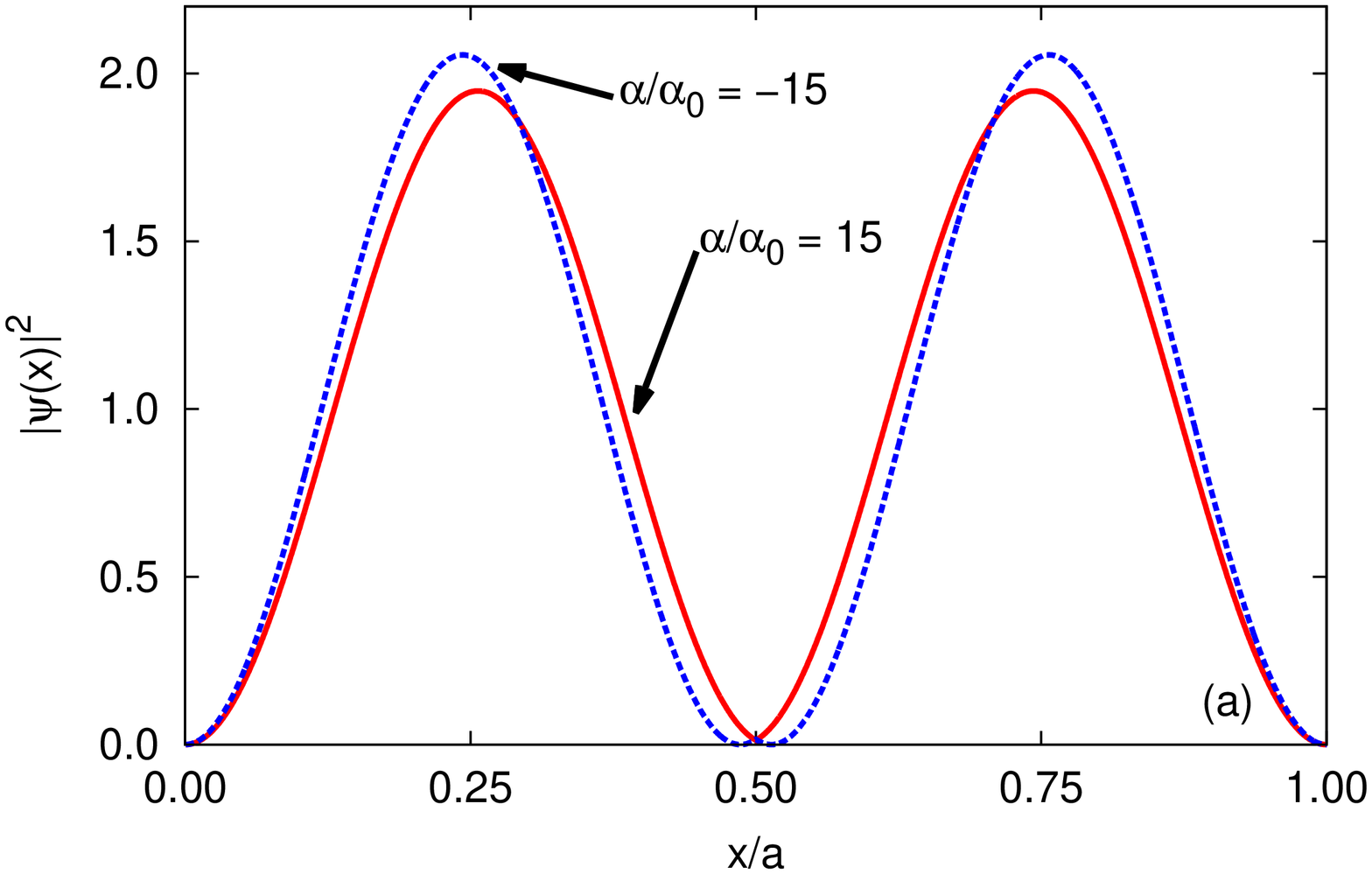}
\includegraphics[width=6.2cm,angle = -90]{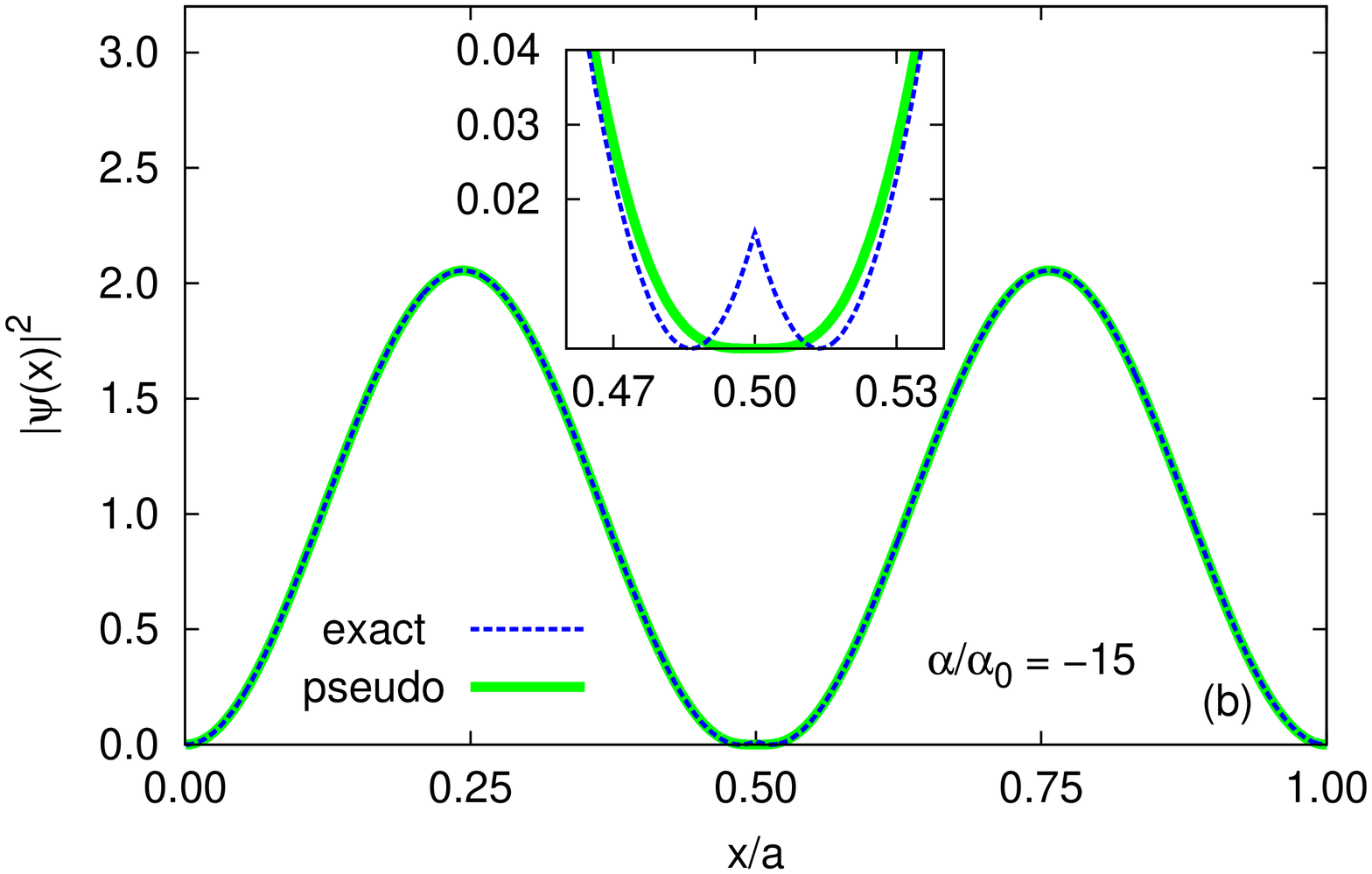}
\caption{(a) Comparison of the probability densities for the first even scattering state for the repulsive $\delta$-function potential
(with $\alpha/\alpha_0 = 15$) (shown with solid red curve) and the attractive $\delta$-function potential  
(with $\alpha/\alpha_0 = -15$) (shown with 
dashed blue curve), vs. position.  Here, $\alpha_0 \equiv a \pi^2 \hbar^2/(2ma^2)$. 
Note that for these figures the infinite square well is now situated at $0 < x < a$ for ease of
computation. Note that the attractive potential (the ``moat'') is effectively a wider barrier than the repulsive wall, as evidenced
by the degree to which the probability density is excluded from the central region. 
In (b) we focus on the moat and show the exact result in comparison with the result obtained from the pseudopotential formulation.
As stated in the text we can transform back and obtain a result in complete agreement with the exact one; instead, following the
historical procedure, we show just the ``smooth'' part. This wave function agrees with the exact result everywhere except in the
``core'' region, where the orthogonalization process allows the wave function to remain smooth in this region, and yet produce
essentially exact eigenvalues.}
\label{Fig4}
\end{figure}
Then Eq.~(\ref{mat1}) is completely defined, and we can proceed to solve this equation numerically. Before showing these solutions,
however, it is worth noting that in the coordinate representation this Schr\"odinger equation becomes
\begin{equation}
-{\hbar^2 \over 2 m} {d^2 \psi(x) \over dx^2} + V(x) \psi(x) + \int \ dx^\prime V_{\rm nl}(x,x^\prime) \psi(x^\prime)  = E\psi(x),
\label{coord}
\end{equation}
which contains two peculiarities: first, it has a non-local potential given by the expression 
$V_{\rm nl}(x,x^\prime) = (E-E_b)\phi_b(x) \phi_b(x^\prime)$, and second,
the non-local potential depends on the eigenvalue $E$ that we are seeking. In the limit of a very strongly bound state, i.e. $\alpha
\rightarrow -\infty$ and therefore $E_b \rightarrow -\infty$ and $\kappa \rightarrow \infty$, the bound state approaches a 
$\delta$-function so the potential induced by the procedure of orthogonalizing to the bound state begins to look like a local potential,
given by $(E-E_b)|\phi_b(x)|^2$. This induced (repulsive) $\delta$-function-like potential readily
overcomes the attractive $\delta$-function, and this forms the effective barrier for the scattering states in the moat problem. 

In Fig.~\ref{Fig4} we show a comparison of the exact result for the probability density 
of the lowest energy even scattering state (as already mentioned, odd states are unaffected by the $\delta$-function
potential), for both the repulsive (wall) and attractive (moat) barriers. Both cases display a characteristic feature 
of wave functions
in a double well potential --- they are both essentially excluded from the central region where the barrier exists (odd states automatically
have this property). Two dissimilar features are striking, and they are related to one another. 
One is the presence of the central cusp for the 
moat potential, indicative that there is some tendency for the probability to increase near the attractive well. The second is that 
the wave function for the moat is in general \textit{more} excluded from the central region, i.e. one could argue that the moat is 
more of a barrier between the two regions than the wall. The reason is that the repulsive region is generated through the bound
state; not only is the bound state more extended than a $\delta$-function, but as indicated above, it gives rise to a non-local
potential, so that its effects will be naturally more extended. The degree to which this is true is clearly depicted in Fig.~\ref{Fig4}~(a),
where the wave function for the moat seems to be the result of a wider repulsive potential barrier.

The second part of this figure shows a comparison of the wave function generated through the pseudopotential with the
exact result. They look very similar, although the insert shows a significant difference near the $\delta$-function potential,
namely, the wave function generated through the pseudopotential remains very smooth through the core region. This characteristic
is very representative of what happens in the orthogonalization process. Here, since we have the wave function for the
bound state we can transform back to the original basis and get essentially exact agreement. This includes structure induced by
the requirement that the scattering wave functions be orthogonal to the bound state. But this transformation is unnecessary, since we
are mainly interested in the eigenvalues, or, even when we require the wave functions, it is the interstitial part (in this case
the part of the wave function away from the $\delta$-function) in which we are interested. As a result we have simply
plotted the untransformed wave function, which is smooth throughout the $\delta$-function region, but agrees very well
with the exact result once one is away from the $\delta$-function.

\begin{figure}[h]
\includegraphics[width=5.0cm,angle = -90]{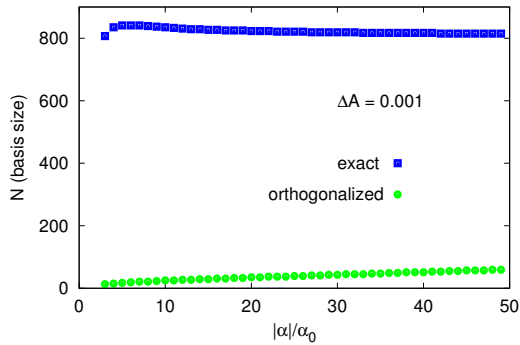}
\caption{With $\Delta A$ as defined in the text set equal to $0.001$, this figure shows the basis size $N$ as a function of
$|\alpha|/\alpha_0$, required to achieve this value of $\Delta A$. Here, $\alpha_0 \equiv a \pi^2 \hbar^2/(2ma^2)$.
For a straightforward diagonalization based on standing wave
states, the required $N$ is shown by the blue squares, while for the expansion in terms of orthogonalized plane wave states
with the approximate pseudopotential, the required $N$ is shown by the green circles. Clearly the same accuracy is achieved
with far fewer basis states in the orthogonalized basis state approach.}
\label{Fig5}
\end{figure}
Recall that one of the motivations for introducing such a procedure was to improve the efficiency of the numerical method
required to solve such problems in general.
That is, convergence as a function of the number of plane wave states (or, in this case, of standing wave states) is expected 
to be more rapid when orthogonalization is used. To illustrate this we have constructed a dimensionless figure of merit, which is
\begin{equation}
\Delta A \equiv \int \ dx \ \left| \left| \psi_{\rm ex}(x) \right|^2 - \left| \psi_{\rm N}(x) \right|^2 \right|,
\label{merit}
\end{equation}
where the subscript $N$ denotes that the wave function was determined through diagonalization with $N$ basis states,
i.e. one requires matrix diagonalization of an $N\times N$ matrix. This measure represents the sum total of the differences
in the probability densities of the two wave functions. Therefore, as the solution becomes more accurate (with increasing $N$),
$\Delta A$ becomes smaller. The point is that for a given tolerance (some choice
of $\Delta A \approx 10^{-3}$ or $10^{-4}$), we anticipate that a smaller matrix size is required if we use the orthogonalization
procedure. With an error of $10^{-3}$ we can barely tell the plotted probability density as a function of $x$ from the
exact solution, so we proceed with this value.
We show results in Fig.~\ref{Fig5} comparing direct diagonalization 
with that using the 
orthogonalization process discussed above. It is clear from this figure that using orthogonalization achieves the same
accuracy much more efficiently.

\vskip0.5cm

\phantom{aaa}

\section{Summary}

We have presented and examined the properties of what we consider an interesting system --- a double well potential
whose barrier consists of a physical moat, rather than a positive potential ``wall'' as is normally the case. A straightforward
solution illustrates that this system shares similar ``double well'' properties with a normal double well. The reason it does so
is because scattering states are required to be orthogonal to the bound states. It is important to realize that this is true
regardless of whether the bound state is occupied or not. Therefore, as far as the scattering states are concerned, the
effective barrier exists even if the bound states are unoccupied, i.e. the argument here does not involve Coulomb repulsion,
or even the Pauli exclusion principle. 

We have also provided a simple demonstration of the orthogonalization procedure. Because of the simplicity of the model
studied, we were also able to construct an explicit pseudopotential and solve this problem as well. Aside from being an interesting
problem in its own right, this example also serves as a ``conduit'' to more realistic problems pertinent to electrons in solids.
We were also able to demonstrate the practical improvement in efficiency that occurs with pseudopotential methods.

\begin{acknowledgments}

This work was supported in part by the Natural Sciences and Engineering Research Council of Canada (NSERC) and by the
Department of Physics at the University of Alberta. We also which to acknowledge a University of Alberta Teaching and Learning
Enhancement Fund (TLEF) grant received a number of years ago that in fact stimulated this work. We also thank Don Page
and Joseph Maciejko for discussions and interest in this problem. 

\end{acknowledgments}

\end{document}